\documentclass[12pt,doublespace]{article}
\begin{document}
\begin{center}
SYNCHRONIZATION IN MULTIPLE TIME DELAY CHAOTIC LASER DIODES SUBJECT TO INCOHERENT OPTICAL FEEDBACKS AND INCOHERENT OPTICAL INJECTION\\
E.M.Shahverdiev $^{1,2}$ and K.A.Shore $^{1}$\\
$^{1}$School of Electronic Engineering,Bangor University,Dean St.,Bangor, LL57 1UT, Wales, UK\\
$^{2}$Institute of Physics, H.Javid Avenue,33, Baku, AZ1143, Azerbaijan\\
ABSTRACT
\end{center}
We present the first report of the synchronization regimes in both unidirectionally and bidirectionally-coupled multiple time delay  chaotic laser diodes subject to incoherent optical feedbacks and incoherent optical injection. 
We derive the existence conditions and numerically study the stability  for lag, complete, and anticipating  synchronization regimes and cascaded synchronization. We also study in detail the effect of parameter mismatches and noise on the synchronization quality. It is emphasized that sensitivity of the synchronization quality to parameter mismatches can lead to a high level of security due to the difficulty to replicate the receiver laser.
We show that the injection current and feedback delay times are highly important parameters from this point of view.\\
~\\
PACS number(s):05.45.Xt, 42.79.-e, 05.45.Vx, 42.55.Px, 42.65.Sf\\
\begin{center}
INTRODUCTION
\end{center}
\hspace*{0.7cm} Chaos synchronization [1] is one of the main features in nonlinear science and plays an important role in a variety of complex physical, chemical, and biological systems, see e.g.references in [2]. Particularly, chaos synchronization has been intensively studied in various nonlinear dynamical systems for its potential applications in chaotic communications, information processing, nonlinear systems performance optimization,etc.In a chaos based secure communications a message is masked in the broadband chaotic output of the transmitter laser and 
synchronization between the transmitter and receiver systems is used to recover a transmitted message [2-5]. \\
Semiconductor lasers find very wide applications e.g. from CD and DVD players to optical communications networks. 
In some of these applications light often re-enters the laser after reflection at other optical elements.Due to their widespread availability and the ease
with which they may be operated in a chaotic regime ,for example, by using external-cavity feedback  [2-5], semiconductor lasers have also been
extensively studied with a view to their utilisation in chaos-based communication systems.\\
\indent Delay differential equations(DDEs) are a large and important class of dynamical systems. These  equations  arise in many scientific and
engineering areas such as optics, economics, networks where they model effects arising due to the finite propagation velocity of information, from the
latency of feedback loops and so on [6]. The behaviour of semiconductor lasers subject to optical feedback(s) (whether delibrate or unwanted) can be
described by a set of DDEs. In many practical applications it may occur that such lasers are subject to more than one optical reflection and thus may
represent  a system with multiple time delays. In addition to modelling such behaviour in semconductor lasers, DDEs with multiple time-delays may provide
more realistic models of interacting complex systems. From the applications viewpoint additional time delays offer the opportunity e.g. to stabilize the
output of a nonlinear system [7]. It would be of great significance also to investigate the influence of additional time delays on the synchronization quality between multiple time delay systems.\\
\indent  When coherent optical feedback is applied to lasers for use in chaos-based communication systems, the chaos synchronization quality depends
sensitively on the frequency detuning between the transmitter and receiver lasers. In particular detuning by a few hundred megahertz between the lasers
would be expected  to effect a significant degradation of the synchronization quality and in turn will impact the efficiency with which
message extraction at the receiver laser can be accomplished. For practical deployment of such chaos-based communications systems it is of great
interest to utilize laser systems where fine tuning of the lasers frequencies is not required, i.e. laser systems where the coherent effect of the phase 
may be neglected. Laser diodes subject to incoherent optical feedback and incoherent injection offer such a facility. In this scheme, the feedback and 
injected fields act on the carrier population in the laser diode active layer but do not interact with the intracavity lasing fields. As a consequence, the phases of the feedback and injection fields do not impact the laser dynamics and synchronization requires no fine tuning of the laser optical frequencies [8-9]. The models used in [8] provide the platform for generalising to the case of multiple incoherent feedbacks and injection considered in the present paper.\\
In this connection, one has to keep in mind that the use of chaotic carrier signals generated in semiconductor lasers using coherent optical feedback also requires coherent optical injection into the receiver laser system to achieve synchronization, however it is very difficult to guarantee such a coherent coupling into the receiver laser after transmission of a chaotic carrier over a long distance [10]. In fact, propagation of the transmitter laser output over distances exceeding the coherence length of the laser will render the laser field incoherent. In this respect it is noted that in the unidrectionally coupled time-delayed systems for complete synchronization to occur the  distance between the lasers should equal the feedback distance. Therefore, the realization of high-speed synchronized chaos, which does not depend on coherent injection and feedback is highly desirable for practical applications.\\
In that case the question of practical realization of synchronization schemes based on incoherent feedback and incoherent injection can be addressed in several ways. In principle, a scheme described in figure 1(a) could be one of the options. In fig. 1(a), which illustrates  a double feedback  scheme the transmitter laser output propagates to two distant mirrors, and then  one part of the reflected beam is injected into the receiver laser and the other one is fed back into the transmitter laser. As the propagation distance of the laser output can be made greater than a coherence length of the laser, then 
a synchronization scheme based on the incoherent feedbacks and incoherent injection is obtained. \\ 
As a another possibility, in the scheme proposed in [8] the linearly polarized output field of the laser first undergoes a $90^{\circ}$ polarization rotation through an external cavity formed by a Faraday Rotator and mirror. It is then split by a non-polarizing beam splitter. One part is fed back into the transmitter laser and the other part is injected into the receiver laser. As the polarization directions of the feedback and injection fields are orthogonal to those of the transmitter and reciever output, respectively, then the transmitter laser is subjected to incoherent optical feedback, while the receiver laser is subjected to incoherent optical injection.  However, as  emphasized in [11-12 ] the simple polarization rotation scheme usually gives rise to both incoherent and coherent optical feedbacks. This fact is also confirmed in recent experimental and theoretical work [13], which considers as an alternative a full two-polarization model. In the case of one short cavity and one or more long cavities the use of a Faraday Rotator in the short cavity arm (fig.1(a)) will be enough also to consider the scheme as one with incoherent optical feedbacks and optical injection.\\ 
\indent A scheme based on the optoelectronic feedback and coupling is another phase-insensitive approach [14-16], see, fig. 1(b).The output from each laser is split by  beamsplitters and directed along different feedback loops and coupling loops. Each signal is converted into  an electronic signal by a photodetector and then amplified before being added to the injection current of a laser. An optical isolator ensures unidirectionality of the coupling. It is noted that in this scheme  that in order to eliminate interference effects between laser fields their detuning should excced the detection bandwidth. As emphasized in [16], apart from avoiding the complexity introduced by the phase of the electric field,  there is another advantage in using the optoelectronic feedback and coupling scheme. Namely in this case there is no restriction to weak or moderate coupling and feedback to avoid secondary round trips in the external cavity. At the same time one must take into account that at very large coupling strengths saturation effects in photodetectors and amplifiers may become significant. Also, in the optoelectronic case the bandwidth of the electronics may act as a low-pass filter on the full dynamics of the optical field [12].\\
It should be emphasized that mathematically in the cases of both incoherent feedback-incoherent injection and optoelectronic feedback-optoelectronic injection  one deals with  similar rate equations for the photon density and the carrier density. Moreover, an all-optical incoherent feedback and incoherent injection system is dynamically equivalent to the optoelectronic system [16]. Equivalence between the dynamics of laser diodes with incoherent optical feedback and injection and dynamics of laser diodes with optoelectronic feedback and coupling is also underlined in recent work [13].\\
\indent As synchronization between the transmitter and receiver lasers is vital for message decoding in chaos-based secure communications, it is 
of paramount importance to investigate chaos synchronization regimes in multiple time delay laser systems. In this paper we present the first report 
of chaos synchronization regimes between laser diodes with multiple incoherent feedbacks and incoherent injection. We derive existence conditions for lag,complete, and anticipating synchronization regimes and numerically study the stability of the synchronization regimes.\\
\indent It can be envisaged that in chaos-based communication systems there will be a need to broadcast a message to a number of receivers or else use may be made of repeater stations to extend the transmission distance. Both cases rely on the possibility of synchronization between a master laser and a number of slave lasers. In other words the study of cascaded synchronization is of great practical importance. In the light of this we also study cascaded synchronization of laser systems. We also present the results of a detailed investigation on the effect of parameter mismatches and noise on the synchronization quality.\\
\begin{center}
SYSTEM MODELLING
\end{center}
\hspace*{0.7cm} Throughout the paper we deal with laser systems with double delay times. To model the arrangement we generalise the approach of [8] to the case of multiple feedback lasers. Thus the dynamics of the double delay time master laser is governed by the following system of equations:
\begin{equation}
\frac{dP_{1}}{dt}=(G_{1}-\frac{1}{\tau_{p1}})P_{1} + \beta_{1} N_{1}
\end{equation}
\begin{equation}
\frac{dN_{1}}{dt}=I_{1} - \frac{N_{1}}{\tau_{s1}} - G_{1}(P_{1} + k_{1} P_{1}(t-\tau_{1}) + k_{2} P_{1}(t-\tau_{2}))
\end{equation}
The receiver laser is described by the following set of equations
\begin{equation}
\frac{dP_{2}}{dt}=(G_{2}-\frac{1}{\tau_{p2}})P_{2} + \beta_{2} N_{2}
\end{equation}
\begin{equation}
\frac{dN_{2}}{dt}=I_{2} - \frac{N_{2}}{\tau_{s2}} -G_{2}(P_{2} + k_{3} P_{2}(t-\tau_{1}) + k_{4} P_{2}(t-\tau_{2}) + KP_{1}(t-\tau_{3})) 
\end{equation}
where $G_{j}=G_{Nj}(1-\epsilon_{j}P_{j})(N_{j}-N_{0j})$, with $j=1$ for the transmitter and $j=2$ for the receiver.
In Eqs.(1-4), $P_{j}$ and $N_{j}$ are the photon number and the electron-hole pair number, respectively, in the
active region of laser $j$.$N_{0j}$ is the value of $N_{j}$ at transparency.$\tau_{pj},\tau_{sj},I_{j},G_{Nj},$
and $\epsilon_{j}$ are respectively the photon lifetime, the carrier lifetime,the injection current (in units of the electron charge),
the gain coefficient, and the gain saturation coefficient of laser $j.$ $\beta_{1}$ and $\beta_{2}$ are the spontaneous emission rates.
$k_{1,2}$ and $k_{3,4}$ are the feedback rates for the transmitter and receiver systems,respectively. $\tau_{1,2}$
are the feedback delay times in the transmitter and receiver systems;$K$ is the coupling rate between the
transmitter and the receiver;$\tau_{3}$ is the time of flight between lasers. Unless otherwise stated, the parameters of the lasers are chosen to be identical, except for the feedback levels and coupling strengths. Throughout this paper $x_{\tau} \equiv x(t-\tau).$\\
Now comparing
\begin{equation}
\frac{dP_{1,\tau_{3}-\tau_{1}}}{dt}=(G_{1,\tau_{3}-\tau_{1}}-\frac{1}{\tau_{p1}})P_{1,\tau_{3}-\tau_{1}},
\end{equation}
\begin{equation}
\frac{dN_{1,\tau_{3}-\tau_{1}}}{dt}=I_{1} - \frac{N_{1,\tau_{3}-\tau_{1}}}{\tau_{s1}} - G_{1,\tau_{3}-\tau_{1}}(P_{1,\tau_{3}-\tau_{1}} + k_{1} P_{1,\tau_{3}} + k_{2} P_{1,\tau_{2}+\tau_{3}-\tau_{1}}) 
\end{equation}
with the receiver system, Eqs.(3-4) one finds that
\begin{equation}
P_{2}=P_{1,\tau_{3}-\tau_{1}},N_{2}=N_{1,\tau_{3}-\tau_{1}}
\end{equation}
is the synchronization manifold under the existence conditions
\begin{equation}
k_{1}=k_{3} + K, k_{2}=k_{4}.
\end{equation}
Analogously we find that
\begin{equation}
P_{2}=P_{1,\tau_{3}-\tau_{2}},N_{2}=N_{1,\tau_{3}-\tau_{2}}
\end{equation}
is the synchronization manifold if
\begin{equation}
k_{2}=k_{4} + K, k_{1}=k_{3}.
\end{equation}
We notice that depending on the relation between the coupling $\tau_{3}$ and
feedback $\tau_{1}$ delay times, manifold (7) is the retarded ($\tau_{3}>\tau_{1}$),
complete ($\tau_{3}=\tau_{1}$),and anticipating ($\tau_{3}<\tau_{1}$) synchronization manifold [17], respectively. It is also noted that with additional time delay the number of possible synchronization manifolds is doubled. Note that existence conditions obtained here are similar to those derived in [18] for semiconductor lasers with coherent feedbacks and injection, i.e.these conditions are quite generic.\\
\indent Most chaos based communication techniques use synchronization in unidirectional master-slave system.Such a coupling scheme prevents the messages 
being exchanged between the sender and receiver. A two way transmission of signals requires bidirectional coupling. With this in mind in 
the paper we also consider chaos synchronization between bidirectionally coupled laser diodes with double time delays. For the bidirectional coupling the systems to be synchronized are described by the equations:
\begin{equation}
\frac{dP_{1}}{dt}=(G_{1}-\frac{1}{\tau_{p1}})P_{1} + \beta_{1} N_{1}
\end{equation}
\begin{equation}
\frac{dN_{1}}{dt}=I_{1} - \frac{N_{1}}{\tau_{s1}} - G_{1}(P_{1} + k_{1} P_{1,\tau_{1}} + k_{2} P_{1,\tau_{2}}+ K_{1}P_{2,\tau_{3}})
\end{equation}
and
\begin{equation}
\frac{dP_{2}}{dt}=(G_{2}-\frac{1}{\tau_{p2}})P_{2} + \beta_{2} N_{2}
\end{equation}
\begin{equation}
\frac{dN_{2}}{dt}=I_{2} - \frac{N_{2}}{\tau_{s2}} -G_{2}(P_{2} + k_{3} P_{2,\tau_{1}} + k_{4} P_{2,\tau_{2}} + K_{2}P_{1,\tau_{3}}) 
\end{equation}
Now comparing the system of Eqs.(11-12) with the system of Eqs.(13-14) we establish that complete synchronization $P_{1}=P_{2}$ is  possible under the conditions $k_{1}=k_{3},k_{2}=k_{4}, K_{1}=K_{2}.$\\
We note that the existence conditions are necessary for synchronization, but these conditions indicate nothing about the stability of the synchronization manifolds. Due to the high complexity of the model the question of stability of the synchronization regimes cannot be studied by analytical means. To identify stable synchronous regimes use is made of numerical modelling. 
\begin{center}
CASCADED CHAOS SYNCHRONIZATION
\end{center}
\indent In this Section we consider the case of cascaded synchronization between unidirectionally coupled chaotic laser systems as illustrated schematically in fig.2. In this case that the output of the slave system of Eqs. (3) and (4) is injected into the another slave system whose dynamics is given by 
\begin{equation}
\frac{dP_{3}}{dt}=(G_{3}-\frac{1}{\tau_{p3}})P_{3} + \beta_{3} N_{3}
\end{equation}
\begin{equation}
\frac{dN_{3}}{dt}=I_{3} - \frac{N_{3}}{\tau_{s3}} -G_{3}(P_{3} + k_{5} P_{3,\tau_{1}} + k_{6} P_{3,\tau_{2}} + K_{1}P_{2,\tau_{3}}) 
\end{equation}
Then using the approach developed in the previous section we establish that 
\begin{equation}
P_{3}=P_{1,2(\tau_{3}-\tau_{1})},N_{3}=N_{1,2(\tau_{3}-\tau_{1})}
\end{equation}
is the synchronization manifold under the existence conditions
\begin{equation}
k_{1}=k_{3} + K, k_{1}=k_{5} + K_{1},k_{2}=k_{4}=k_{6}.
\end{equation}
Analogously we find that
\begin{equation}
P_{3}=P_{1,2(\tau_{3}-\tau_{2})},N_{3}=N_{1,2(\tau_{3}-\tau_{2})}
\end{equation}
is the synchronization manifold if
\begin{equation}
k_{2}=k_{4} + K,  k_{2}=k_{6} + K_{1}, k_{1}=k_{3}=k_{5}.
\end{equation}
We emphasize that existence conditions for $P_{3}=P_{1,2(\tau_{3}-\tau_{1})},N_{3}=N_{1,2(\tau_{3}-\tau_{1})}$ ($P_{3}=P_{1,2(\tau_{3}-\tau_{2})},N_{3}=N_{1,2(\tau_{3}-\tau_{2})}$) include existence conditions for synchronization between systems  of Eqs.(1-2) and (3-4). This is  demonstrated below via numerical modelling of the synchronization between systems of Eqs.(1-2),Eqs.(3-4) and  Eqs.(15-16).\\
We notice that having two slave lasers in the chain give rise to a doubling of the time shift between the synchronized systems to become $2(\tau_{3}-\tau_{1})$( or $2(\tau_{3}-\tau_{2}))$. This result of immense practical importance from the application point of view of anticipating synchronization. It is straightforward to establish that with additional slave lasers in the chain one can further increase the anticipation time. Another important conclusion 
from the obtained results is that, complete synchronization takes place provided the feedback time and time of flight are equal, regardless of the communication distance between the transmitter and receiver. In other words the time of flight is not a fundamental restriction on the communication system performance.\\
It is also noted that work by Voss [19] also deals with the cascaded synchronization, but for systems of ordinary differential equations coupled via time delay or in a single time delay systems with delayed coupling. The systems in this paper are more general in considering multiple time delay systems coupled with time delay. These systems are significantly more complex and also have the virtue of forming rather realistic models of physically realisable systems.\\
\begin{center}
NUMERICAL SIMULATIONS
\end{center}
Numerical simulations were conducted using the DDE23 program [20] in
Matlab 7. We use typical values for the internal parameters of the laser diode:
$\tau_{p1}=2 ps,\tau_{s1}=2 ns,G_{N1}=1\times 10^{4} s^{-1}, N_{01}=1.1\times 10^{8}, I_{1}=5.3\times 1.7\times 10^{17},\beta_{1}=5\times 10^{3}$ and put $\epsilon=0.$
Unless otherwise stated, the parameters of the lasers are identical.\\
The quality of chaos synchronization can be quantified by a cross-correlation coefficient between the outputs of the lasers. We have calculated the cross-correlation coefficient [21] using the formula 
\begin{equation}
C(\Delta t)= \frac{<(x(t) - <x>)(y(t+\Delta t) - <y>)>}{\sqrt{<(x(t) - <x>)^2><(y(t+ \Delta t) - <y>)^2>}},
\end{equation}
where $x$ and $y$ are the outputs of the lasers, respectively;the brackets$<.>$
represent the time average; $\Delta t$ is a time shift between laser outputs. This coefficient indicates the quality of synchronization:C=0 implying no synchronization;C=1 indicating perfect synchronization. In the numerical simulations we used time delay values corresponding to the case of long cavities:i.e. time delays of the order of tens of ns, even hundreds of ns. The conclusions for both cases were similar, but in the latter case the numerical calculations required excessive computing time. For this reason  results of computer simulations for the case of several tens of ns. are presented here. \\
\indent First we consider the case of complete synchronization between unidirectionally coupled master (transmitter), Eqs.(1-2) and slave (receiver), Eqs.(3-4) laser systems. According to our analytical findings complete synchronization 
$P_{1}=P_{2}$ could occur if $k_{1}=k_{3} ,k_{2}=k_{4} + K$
for $\tau_{3}=\tau_{2}.$ Figure 3 demonstrates complete synchronization for
$k_{1}=k_{3}=0.41,k_{2}=0.55,k_{4}=0.15,K =0.40 ,\tau_{1}=30ns, \tau_{2}= \tau_{3}=40ns.$ Fig.4 shows lag synchronization $P_{2}=P_{1,\tau_{3}-\tau_{2}}.$ In this case the lag time is $\tau_{3}-\tau_{2}=10ns.$ Anticipating synchronization manifold $P_{2}=P_{1,\tau_{3}-\tau_{2}}$ 
(fig.5,$\tau_{3}<\tau_{2}$) is also observed in full accordance with the analytic results established above.
Note that the synchronization manifold $P_{2}=P_{1,\tau_{3}-\tau_{1}}$ is 
equivalent to $P_{1}=P_{2,\tau_{1}-\tau_{3}}.$  Numerical simulations of bidirectionally coupled laser diodes with double time delays are presented in 
fig.6, which shows synchronization error dynamics $P_{2}-P_{1}.$\\
So as to allow dynamical transients to expire the lag and anticipating regimes time series begin from non-zero value of the normalized time.The
achievement of cross-correlation coefficents  very close to unity evidences the high-quality synchronization. Also it is noted that for high
intensities the synchronization plots display an almost perfect synchronization between the lasers, but during intensity dropouts (low intensities) 
of the drive the lasers desynchronize, see e.g.,fig. 4(b).\\
Figures 7(a) and 7(b) demonstrate cascaded complete chaos synchronization, Eqs.(1-4) and Eqs.(15-16) for $k_{1}=k_{3}=k_{5}, k_{2}=k_{4}+ K = k_{6}+ K_{1}.$ Figure 7(a) shows synchronization error dynamics $P_{2}-P_{1}$ between the first and second lasers. Synchronization error dynamics $P_{3}-P_{1}$ between the first and last lasers is depicted in fig.7(b). We notice that if complete sychronization  existence conditions between $P_{1}$ and $P_{2}$ are violated, then complete synchronization between $P_{1}$ and $P_{3}$ does not occur, fig.7(c). \\
\indent We notice that additional time delays can create new synchronization regimes. A natural question which arises is how can additional time delays influence the synchronization quality of the synchronization regimes for the single time delay systems. In the light of this we have also investigated the influence of the second feedback delay on the synchronization quality, e.g. the correlation coefficient between the synchronized lasers.  The results of numerical simulations have shown (not presented here) the possibilty of stabilizing role being played by the additional time delays,i.e. additional feedback channels can facilitate high quality synchronization- an important result for information processing. It is noted that similar conclusions were reached in [18] for semiconductor lasers with coherent feedbacks and injection. Also, the role of additional time delays in achieving a homogeneous steady state in coupled chaotic maps was investigated in recent work [22] with the stabilizing role of the additional time delays being emphasized. In other words, the stabilizing role of the additional feedbacks is quite widespread.\\
\indent Next we will study the effect of parameter mismatches on the synchronization quality. It is noted that sensitivity of the synchronization to mismatches of the parameters can lead to a high level of security due to the difficulty to replicate the receiver laser, i.e. sensitivity to parameter mismatches increases the security of encryption [8,23]. However, the internal parameters of the interacting laser diodes unlikely to match exactly even if they are produced from the same wafer. Moreover, the operating parameters cannot be perfectly controlled. In other words, in practical cases, synchronization must therefore occur also for small parameter mismatches.  Most importantly, an investigation of the effect of parameter mismatches on synchronization quality will enable determination of the  most sensitive synchronization parameters. In studying the influence of the parameter mismatches on the synchronization quality emphasis is given to cascaded synchronization, where mismatch effects are more pronounced. In our numerical simulations we allow a 6$\%$ mismatch between parameters.\\
In figure 8 the dependence of the cross-correlation coefficient $C$ between the first and third lasers 1 and 3 on the ratio $k_{1}/k_{3}$ of the feedback strength $k_{1}$ to the feedback strength $k_{3}$  and on the ratio $k_{5}/k_{3}$ of the feedback strength $k_{5}$ to the feedback strength $k_{3}$ in the cascaded configuration (Eqs.(1-4) and Eqs.(15-16)) is presented. \\
The similar trend also holds for the mismatch of the coupling strength, i.e. a mismatch between $K_{1}$ and $k_{2}-k_{6}$,and between $K$ and $k_{2}-k_{4}.$ It is emphasized that for these cases the synchronization quality is quite robust for small parameter mismatches (1-2 $\%$).\\
Figure 9 demonstrates the effect of parameter mismatches between the injection currents for the first and last lasers on the cross-correlation coefficient in the cascaded configuration (Eqs.(1-4) and (Eqs.(15-16)), i.e. it displays  the dependence of $C$ on the ratio $I_{1}/I_{2}$  of the injection current $I_{1}$ for laser 1 to the injection current $I_{2}$ for laser 2 and on the ratio $I_{3}/I_{2}$ of the injection current $I_{3}$ for laser 3 to the injection current $I_{2}$ for laser 2.\\
Figure 10 shows how the quality of synchronization changes with the degree of mismatch between the feedback delay times of the same loop for the lasers 1 and 3 (Eqs.(1-4) and Eqs.(15-16)), i.e. it presents the dependence of $C$ on the ratio $\tau_{1}^{(1)}/\tau_{1}^{(2)}$ of the first feedback loop time $\tau_{1}^{(1)}$ for laser 1 to the first feedback loop time $\tau_{1}^{(2)}$ for laser 2 and on the ratio  $\tau_{1}^{(3)}/\tau_{1}^{(2)}$ of the first feedback loop time $\tau_{1}^{(3)}$ for laser 3 to the first feedback loop time $\tau_{1}^{(2)}$ for laser 2. We have also found the similar trend for the second feedback loop, not presented here.\\
These figures demonstrate high sensitivity of the synchronization quality to the parameter mismatches of the injection current and feedback times.\\
\indent Although we have presented results of numerical simulations for parameter mismatches for the case of cascaded unidirectionally coupled lasers, the conclusions are also valid  both for unidirectionally and bidirectionally coupled lasers,i.e in all the cases studied the synchronization quality is very sensitive to the injection current and feedback time mismatches and is rather robust to mismatches in the feedback and coupling rates. At the same time we emphasize that in all the cases analyzed bidirectionally coupled systems perform better than the unidirectionally coupled lasers. It seems that the two mutually coupled chaotic systems influence the dynamics of each other and can accelerate the synchronization by enhancing coherent moves.\\
\indent Finally we consider the effect of noise on the synchronization quality for both unidirectionally and bidirectionally coupled systems. For this purpose the right-hand side of the equation for the photon density is augmented by a Langevin noise force $F,$ which accounts for stochastic fluctuations arising from stontaneous-emission processes. The Langevin forces satisfy the relations $<F_{j}(t)F_{j}(t')>=2N_{j}P_{j}\beta_{j}\delta(t-t')$. 
Figures 11(a) and 11(b) portray the error dynamics $P_{2}-P_{1}$  and $P_{3}-P_{1}$ for two and three uindirectionally coupled systems, Eqs.(1-4) and Eqs(1-4,15-16), respectively. Figure 11(c) shows the error dynamics $P_{2}-P_{1}$ for the bidirectionally coupled systems, Eqs.(11-14). For comparison, in the cases of noiseless error dynamics the synchronization quality was maximum, i.e. C=1, see,e.g. fig. 3(b) and fig. 6. For all figures the data were averaged over 100 time series. It is noted that in general, as expected, noise degrades the synchronization quality, although its affect in the case of cascaded coupled systems is more pronounced and bidirectionally coupled systems show better performance (in terms of synchronization quality) than the unidirectionally coupled lasers.\\
\begin{center}
CONCLUSIONS
\end{center}
\indent To summarize, we have reported on the chaos synchronization regimes in unidirectionally and bidirectionally coupled multiple time delay laser diodes subject to incoherent optical feedbacks and incoherent optical injection. We have derived existence conditions for the synchronization regimes and numerically verified our findings.\\
We have also investigated cascaded synchronization between multiple time delay laser systems. We have established that provided the feedback time and time of flight are equal, regardless of the communication distance between the transmitter and receiver(s) complete synchronization takes place,i.e.the time of flight is not a fundamental restriction on the communication system performance. We have also shown that if 
the feedback times and time of flight are not equal, then increased anticipation times can be achieved using cascaded configuration. Practical applications of the results are seen in the control of delay-induced instabilites in a wide 
range of non-linear systems are for understanding natural information processing [19].\\
We have also studied in detail the effect of parameter mismatches and noise on the synchronization quality. It is emphasized that sensitivity of the synchronization quality to parameter mismatches can lead to a high level of security due to the difficulty to replicate the receiver laser. We have shown that the injection current and feedback delay times are highly important parameters from the viewpoint of encryption security.\\
The results of the paper provide the basis for the use of lasers diodes with multiple incoherent feedbacks and injection in chaos-based secure high-speed communication systems.\\
{\it Acknowledgements}.-This research was supported by a Marie Curie International Fellowship within the $6^{th}$ European Community 
Framework Programme Contract N MIF1-CT-2006-039927.\\
\newpage
\begin{center}
Figure captions
\end{center}
\noindent FIG.1(a). Schematic experimental arrangement for the synchronization of lasers with double incoherent feedback and incoherent injection.
LD:Laser diode.BS:Beamsplitter. M:Mirror. DL:Delay lines. FR:Faraday Rotator. OI:Optical Isolator.\\
~\\
\noindent FIG.1(b). Schematic experimental arangement for the synchronization of lasers with double optoelectronic feedback and injection.LD:Laser diode. PD:Photodetector.BS:Beamsplitter. DL:Delay lines. OI:Optical Isolator. A:Amplifier. The output from each laser is split by a beamsplitters and directed along different feedback loops and coupling loops. Each signal is converted into  an electronic signal by a photodetector and then amplified before being added to the injection current of the laser.\\
~\\
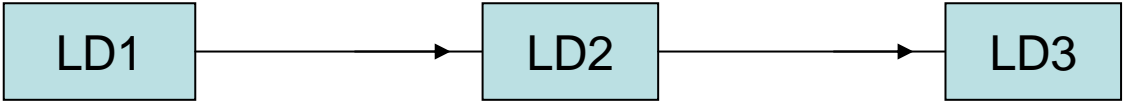
\noindent FIG.2. Schematic diagram of cascaded synchronization between  unidirectionally coupled lasers.\\
~\\
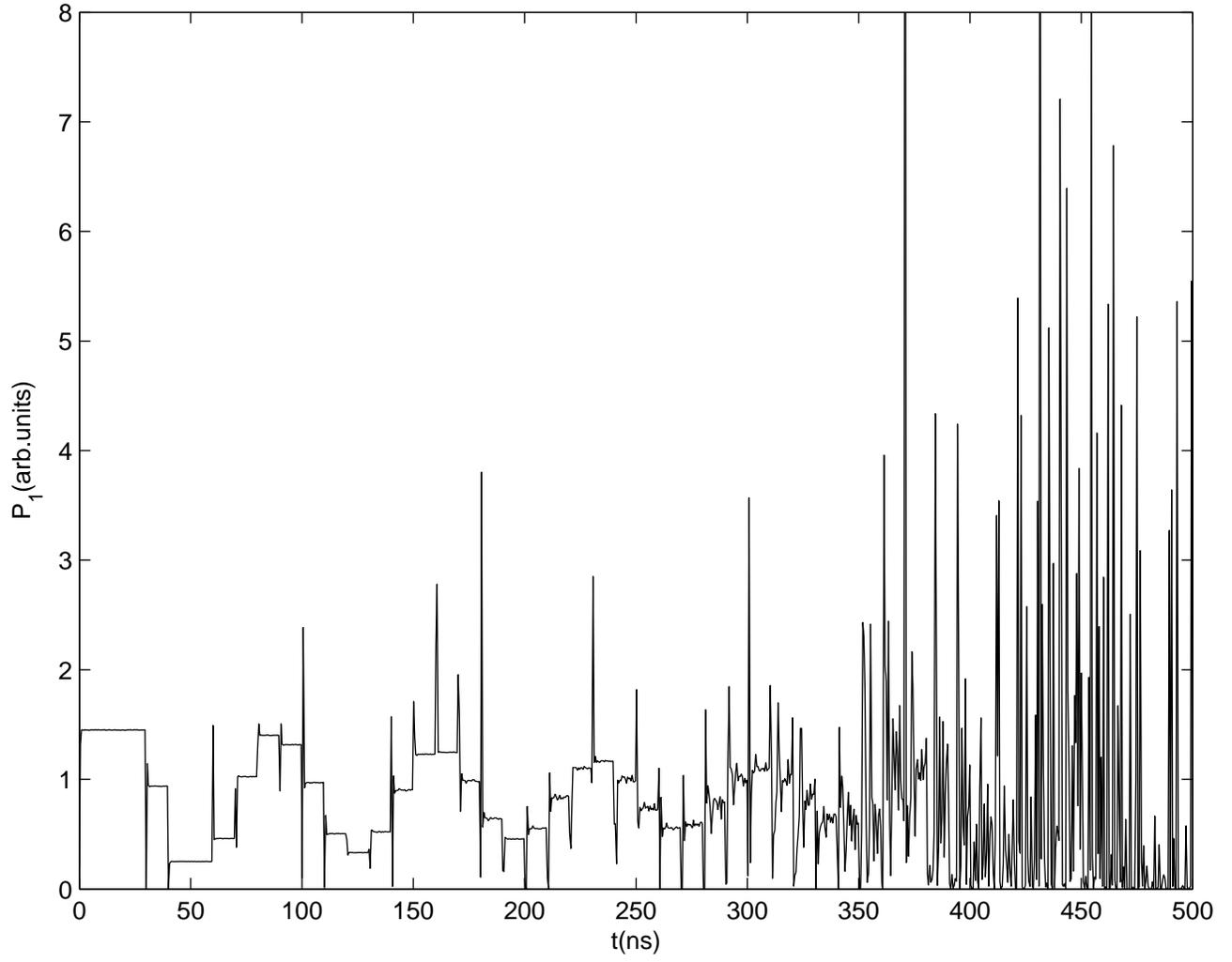
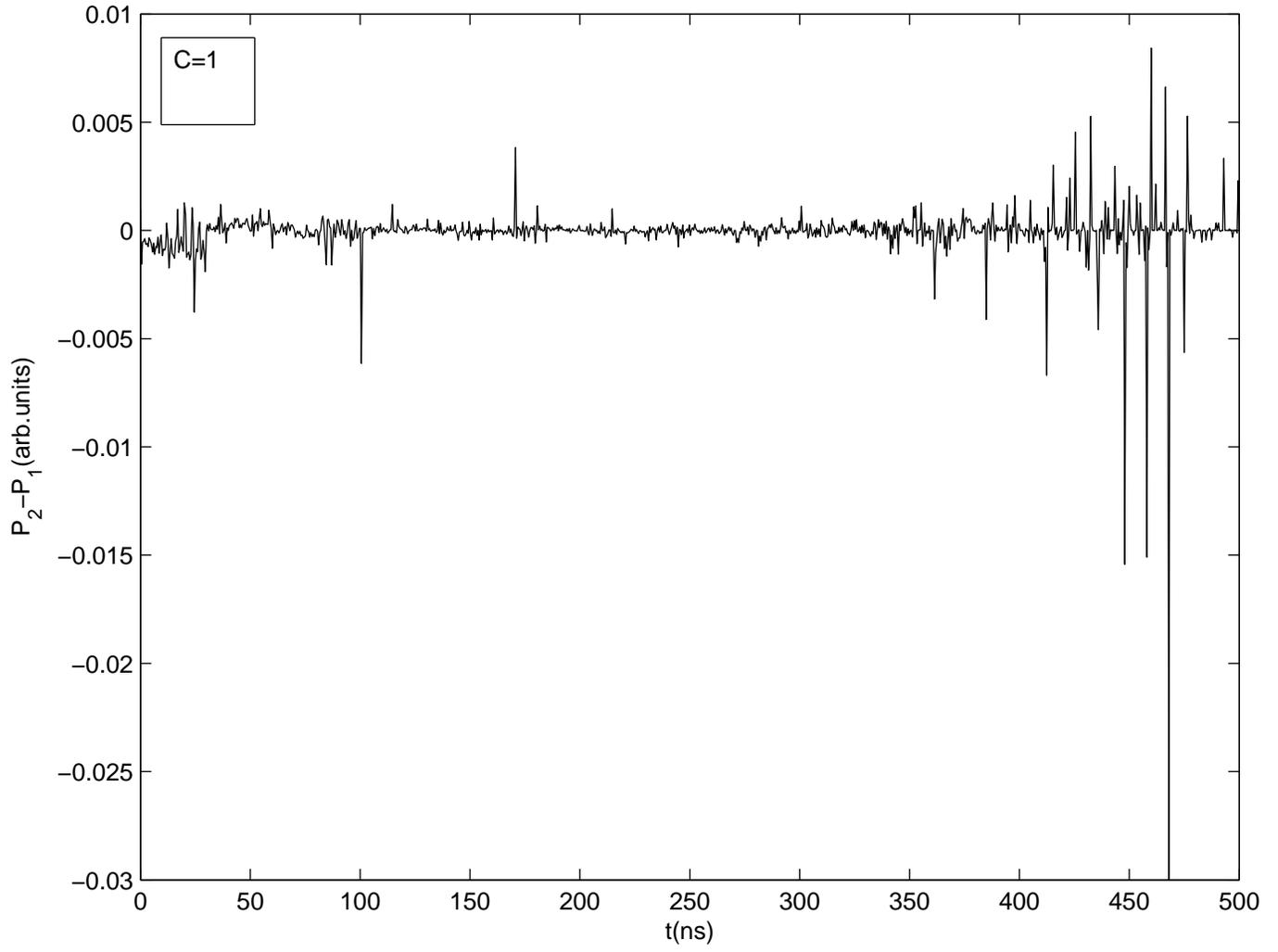
\noindent FIG.3. Numerical simulation of Eqs.(1-4) for
$k_{1}=k_{3}=0.41, k_{2}=0.55, k_{4}=0.15, K =0.40 ,\tau_{1}=30ns, \tau_{2}=\tau_{3}=40ns.$Complete synchronization: (a) time series of the transmitter $P_{1}$ laser;(b) synchronization error $P_{2}-P_{1}$ dynamics.C is the correlation coefficient between $P_{2}$ and $P_{1}.$\\
~\\
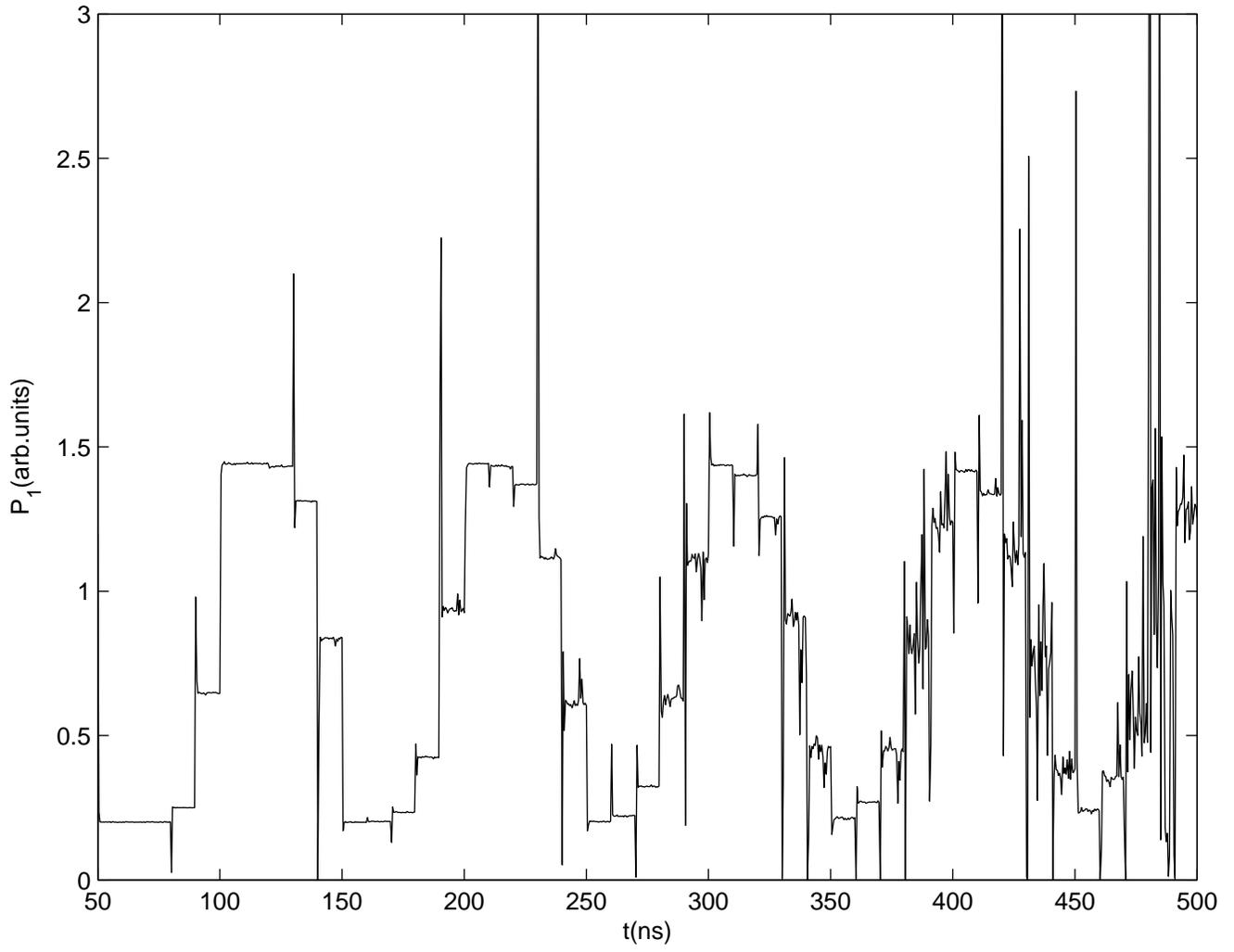
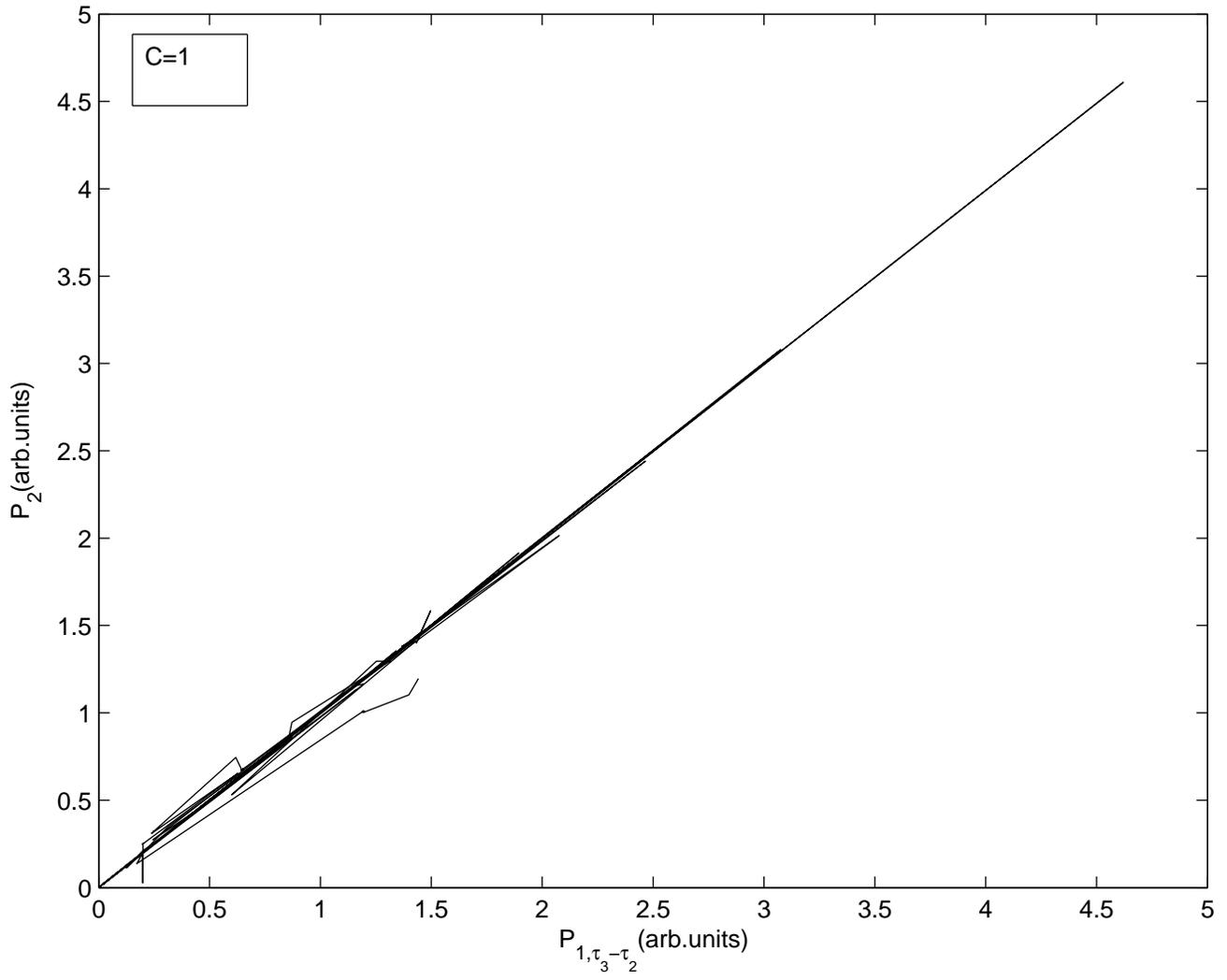
\noindent FIG.4. Numerical simulation of Eqs.(1-4) for $k_{1}=k_{3}=0.20, k_{2}=0.80, k_{4}=0.15, K =0.65 ,\tau_{1}=40ns, \tau_{2}= 50ns, \tau_{3}=60ns.$
Lag synchronization: (a) time series of the transmitter $P_{1}$ laser;(b) correlation plot between $P_{1,\tau_{3}-\tau_{2}}$ and $P_{2}.$C is the correlation coefficient between $P_{1,\tau_{3}-\tau_{2}}$ and $P_{2}.$\\
~\\
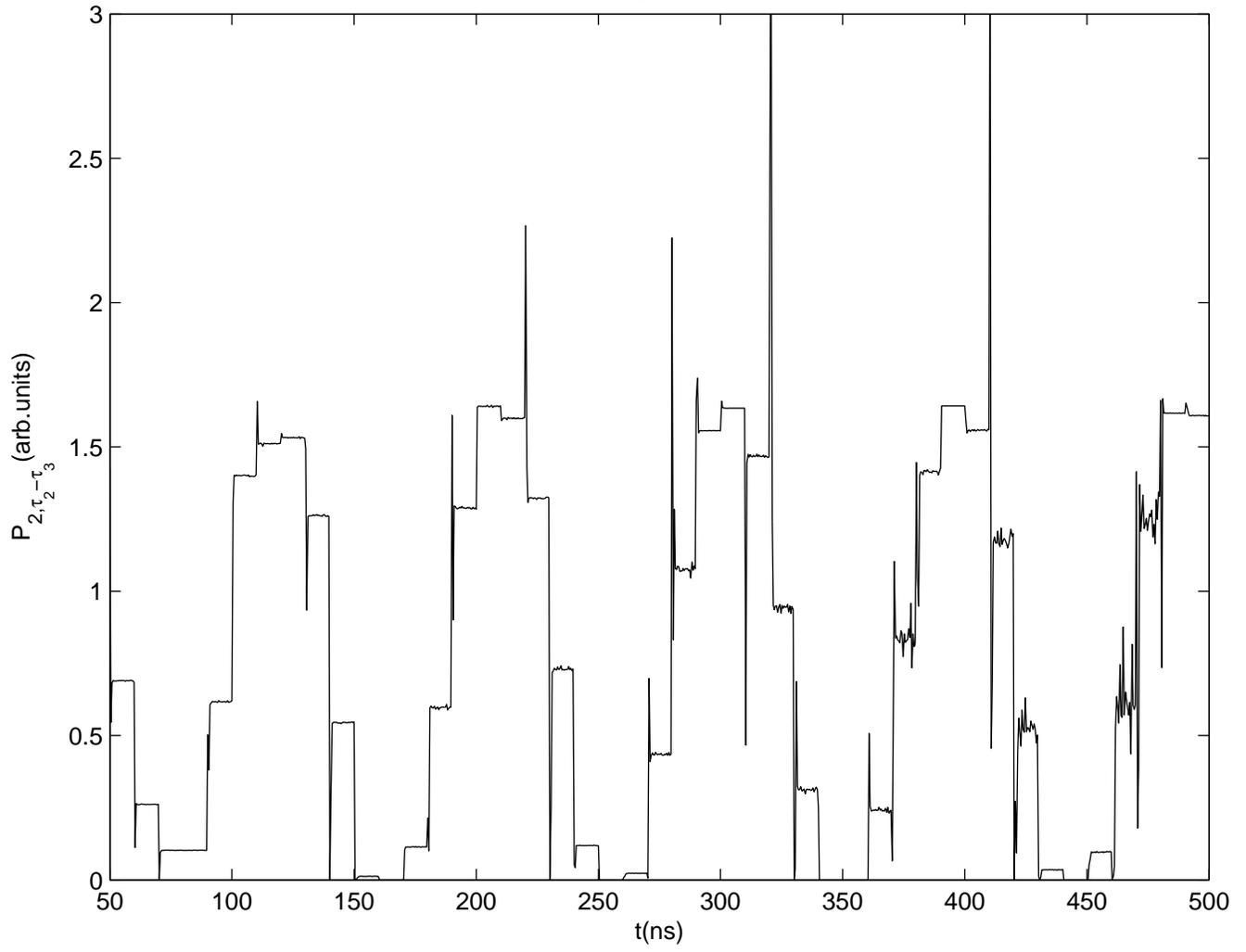
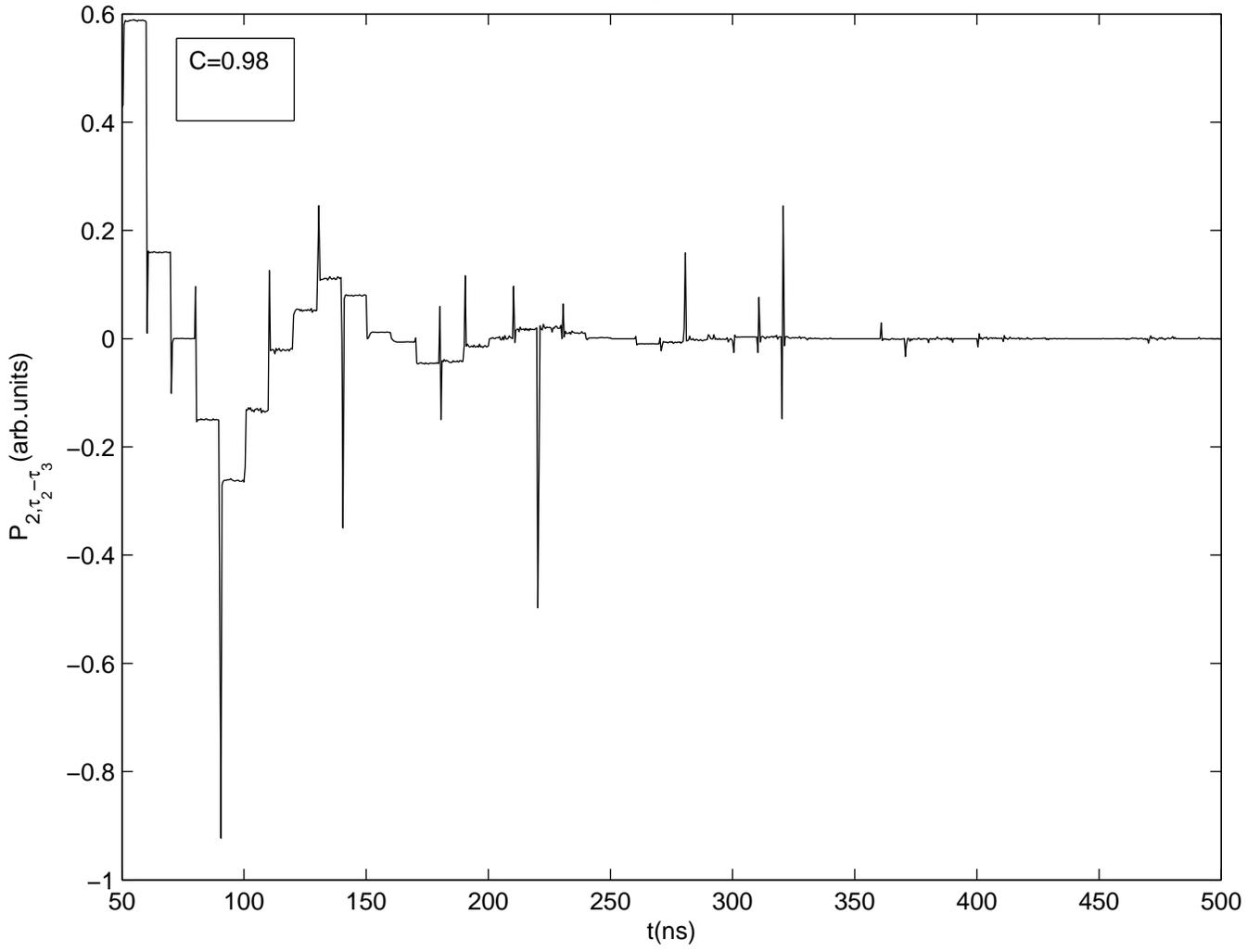
\noindent FIG.5. Numerical simulation of Eqs.(1-4) for $k_{1}=k_{3}=0.35, k_{2}=0.73, k_{4}=0.13, K =0.60 ,\tau_{1}=40ns, \tau_{2}= 50ns, \tau_{3}=30ns.$
Anticipating synchronization : (a) time series of the receiver $P_{2,\tau_{1}-\tau_{3}}$ laser;(b) synchronization error $P_{2,\tau_{1}-\tau_{3}}-P_{1}$ dynamics.C is the correlation coefficient between $P_{2,\tau_{1}-\tau_{3}}$ and $P_{1}.$\\
~\\
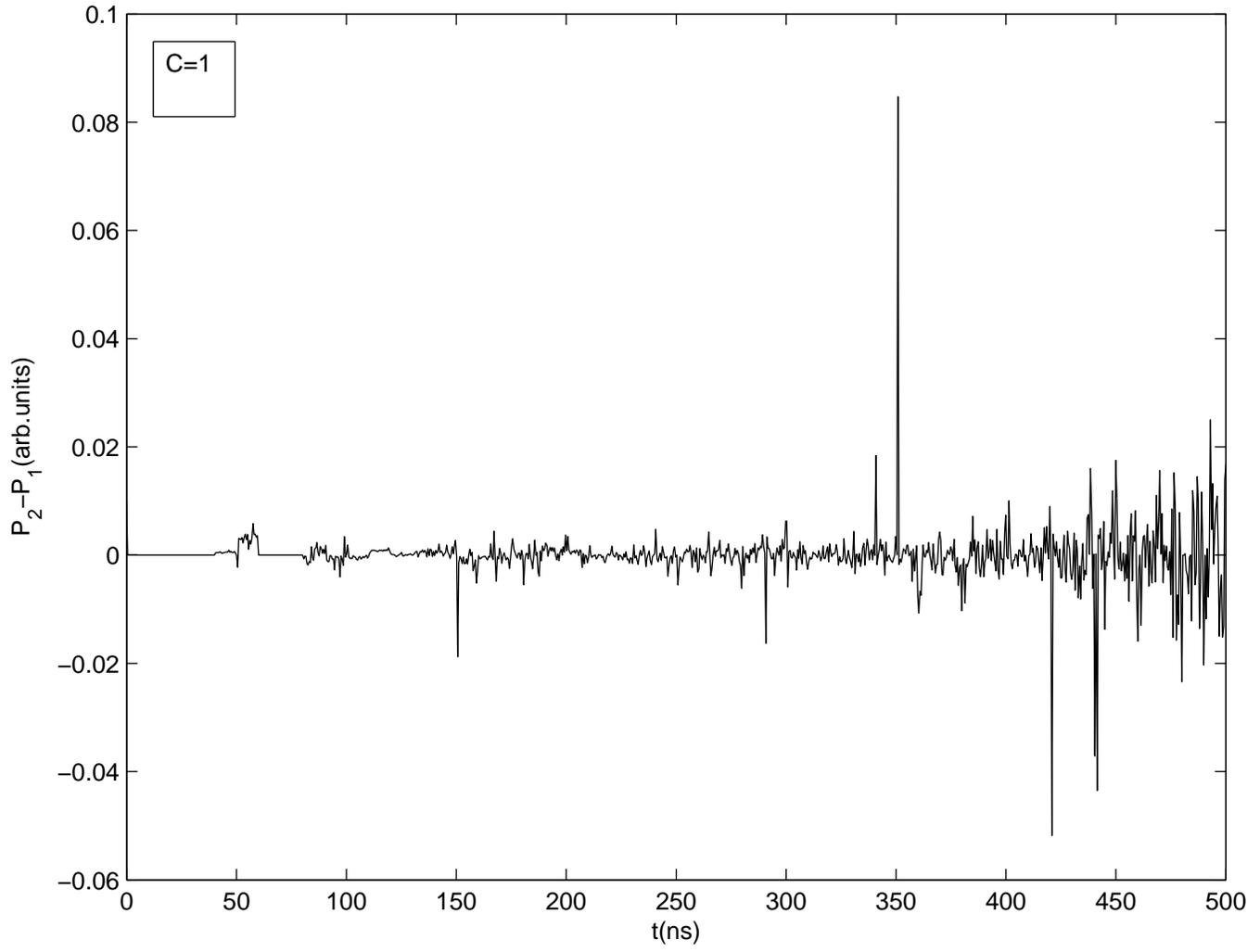
\noindent FIG.6. Numerical simulations of Eqs.(11-14) for bidirectionally coupled lasers for 
$k_{1}=k_{3}=0.41,k_{2}=k_{4}=0.25, K =0.30 ,\tau_{1}= 40ns,\tau_{2}=50ns,\tau_{3}=60ns.$Synchronization error 
$P_{2}-P_{1}$ dynamics.C is the correlation coefficient between $P_{2}$ and $P_{1}.$\\
~\\
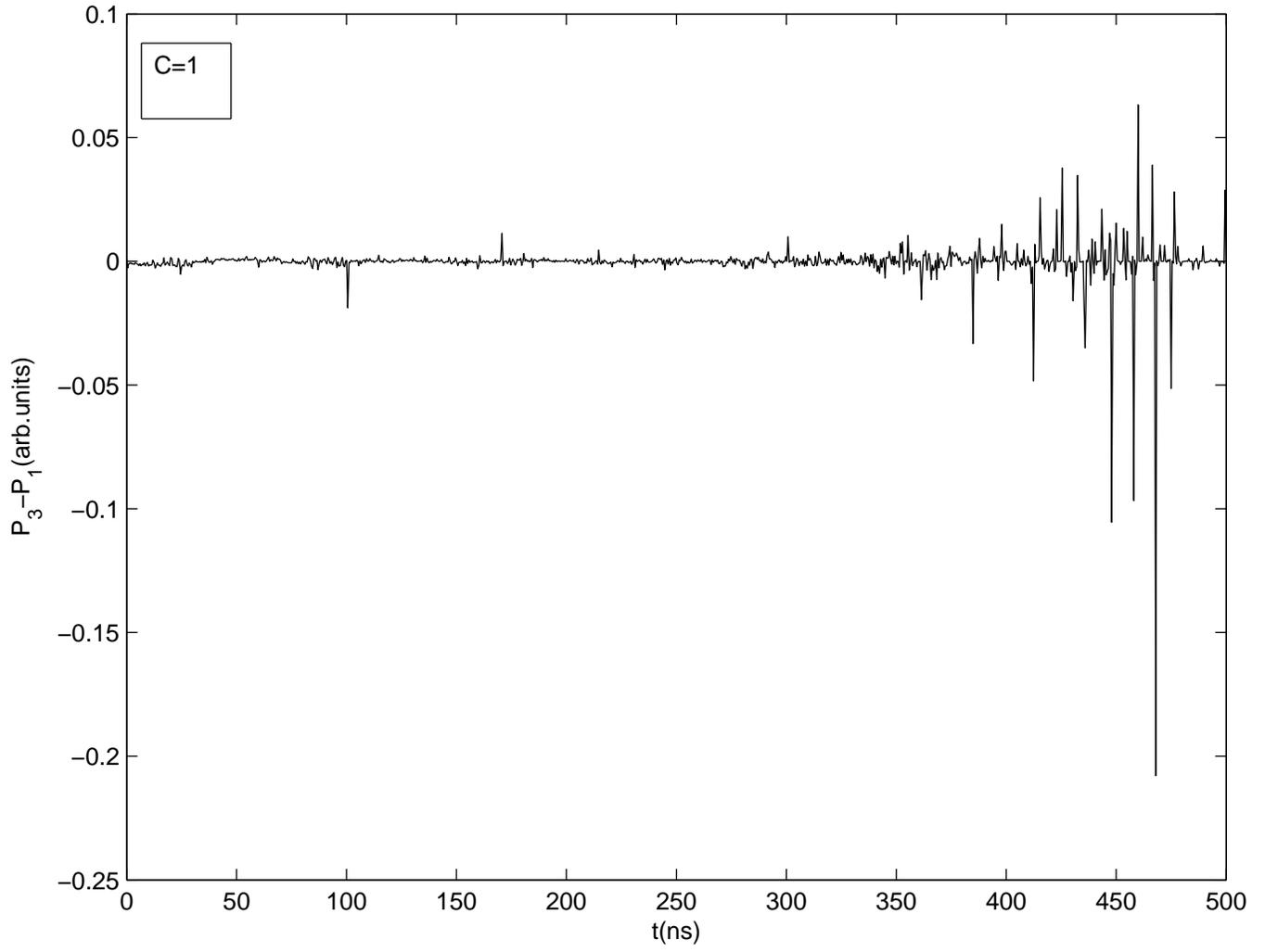
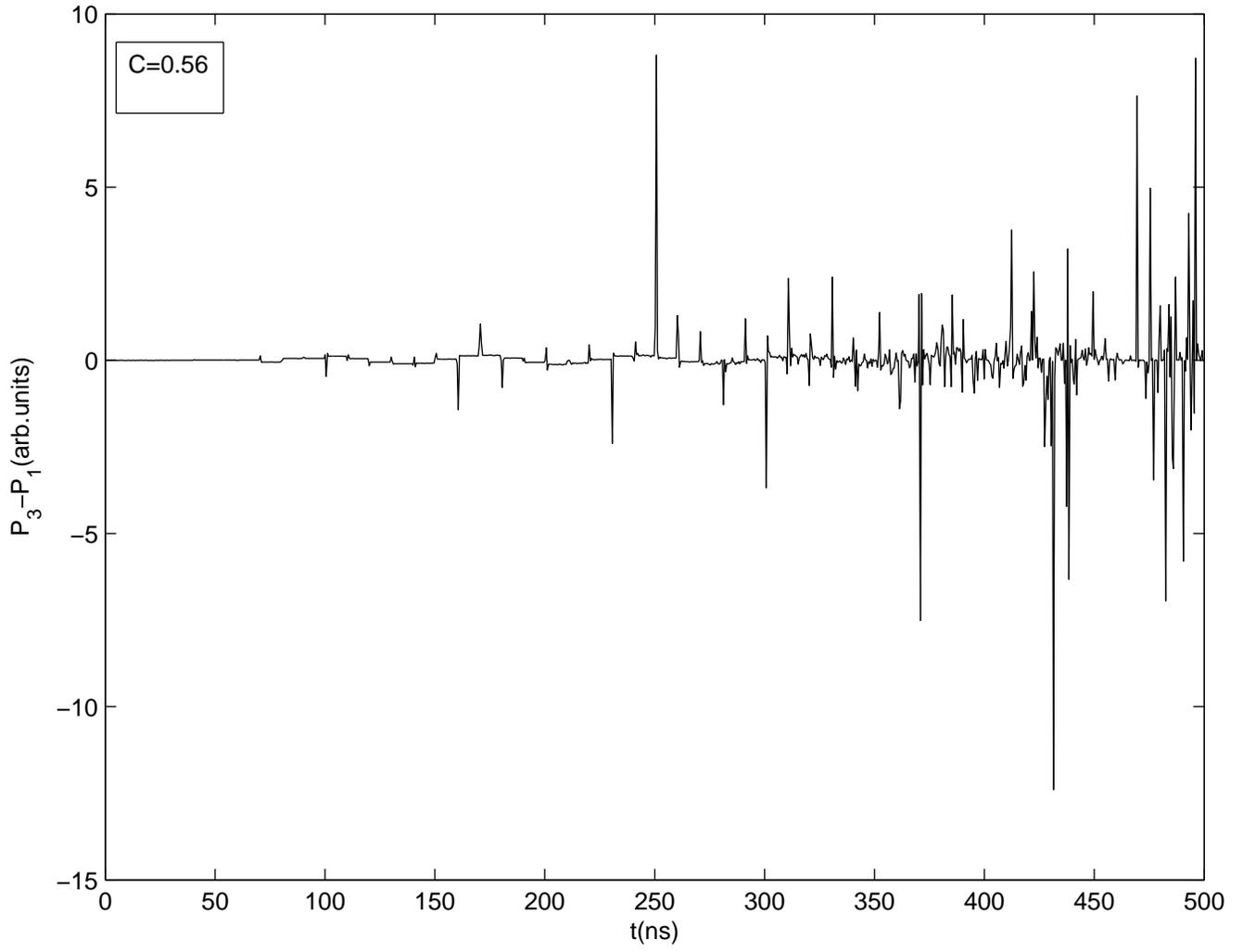
\noindent FIG.7. Numerical simulation of Eqs.(1-4) and (15-16) for
$k_{1}=k_{3}=k_{5}=0.41,k_{2}=0.55, k_{4}=0.15,K =0.40,k_{6}=0.11,K_{1}=0.44,\tau_{1}=30ns, \tau_{2}=\tau_{3}=40ns.$Cascaded complete synchronization:
(b)Synchronization error $P_{3}-P_{1}$ dynamics.C is the correlation coefficient between $P_{3}$ and $P_{1};$
(c)Synchronization error $P_{3}-P_{1}$ dynamics for $k_{3}=0.31,k_{4}=0.25,K=0.50,$ other parameters as in figures 7(a) and 7(b).C is the correlation coefficient between $P_{3}$ and $P_{1}.$\\
~\\
\noindent FIG.8(color). Numerical simulations of Eqs.(1-4) and (15-16) for cascaded unidirectionally coupled lasers for 
$k_{3}=0.41, k_{2}=0.55, k_{4}=0.15, k_{6}=0.11, K=0.40, K_{1}=0.44 ,\tau_{1}= 30ns,\tau_{2}=40ns,\tau_{3}=40ns:$ Dependence of the cross-correlation coefficient $C$ between the first and third lasers 1 and 3 on the ratio of the feedback strength $k_{1}$ to the feedback strength $k_{3}$ and on the ratio of the feedback strength $k_{5}$ to the feedback strength $k_{3}.$ \\
~\\
\noindent FIG.9(color). Numerical simulations of Eqs.(1-4) and (15-16) for cascaded unidirectionally coupled lasers for 
$k_{1}=k_{3}=k_{5}=0.41, k_{2}=0.55, k_{4}=0.15, k_{6}=0.11, K=0.40, K_{1}=0.44 ,\tau_{1}= 30ns,\tau_{2}=40ns,\tau_{3}=40ns:$
Dependence of the cross-correlation coefficient $C$ between the first and third lasers 1 and 3 on the ratio of the injection current $I_{1}$ to the injection current $I_{2}$ and 
on the ratio of the injection current $I_{3}$ to the injection current $I_{2}.$ \\
~\\
\noindent FIG.10(color). Numerical simulations of Eqs.(1-4) and (15-16) for cascaded unidirectionally coupled lasers for 
$k_{1}=k_{3}=k_{5}=0.41, k_{2}=0.55, k_{4}=0.15, k_{6}=0.11, K=0.40, K_{1}=0.44 ,\tau_{1}= 30ns,\tau_{2}=40ns,\tau_{3}=40ns:$
Dependence of the cross-correlation coefficient $C$ between the first and third lasers 1 and 3 on the ratio of the first feedback loop time $\tau_{1}^{(1)}$ for laser 1 to the first feedback loop time $\tau_{1}^{(2)}$ for laser 2 and on the ratio of the first feedback loop time $\tau_{1}^{(3)}$ for laser 3 to the first feedback loop time $\tau_{1}^{(2)}$ for laser 2.\\
~\\
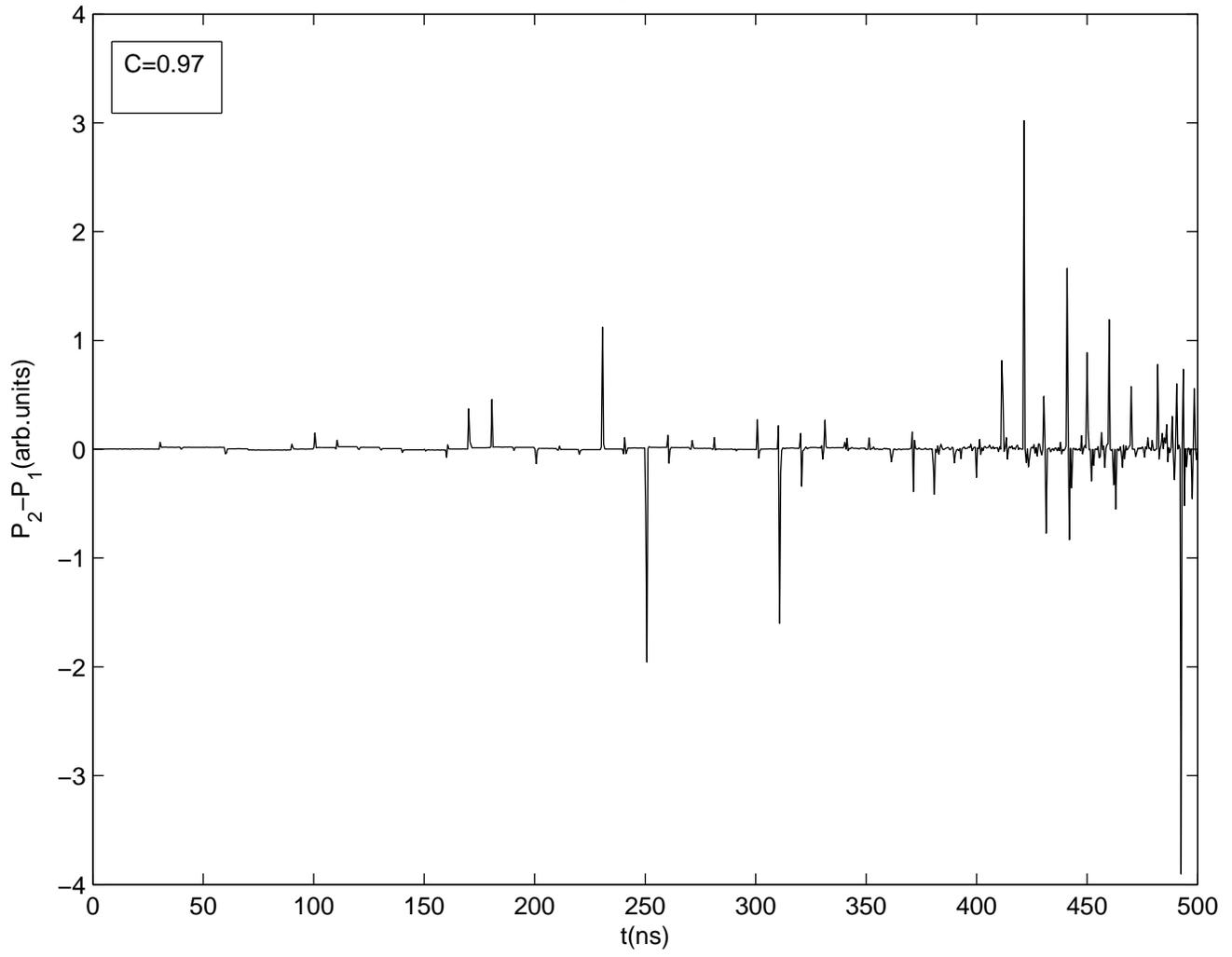
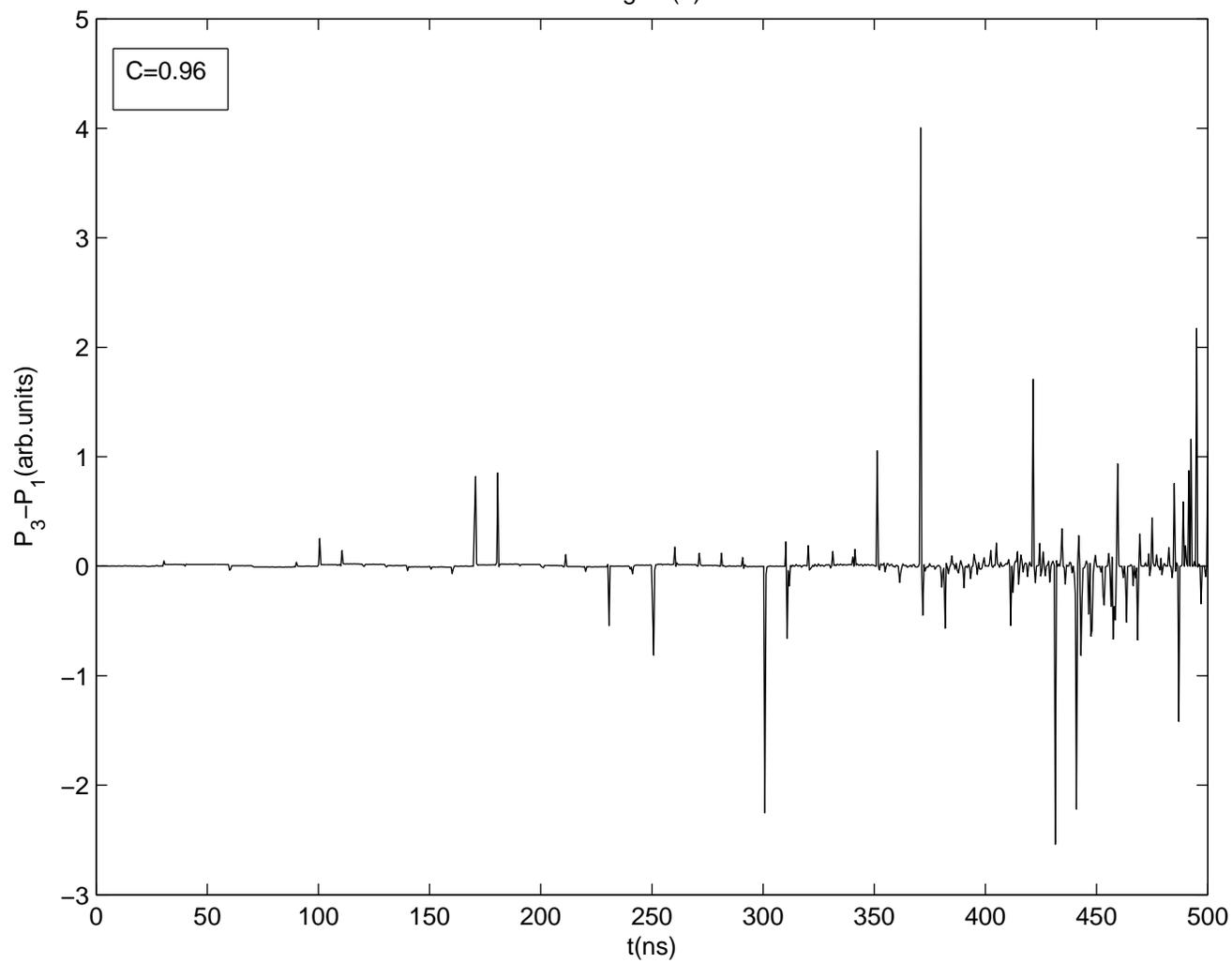
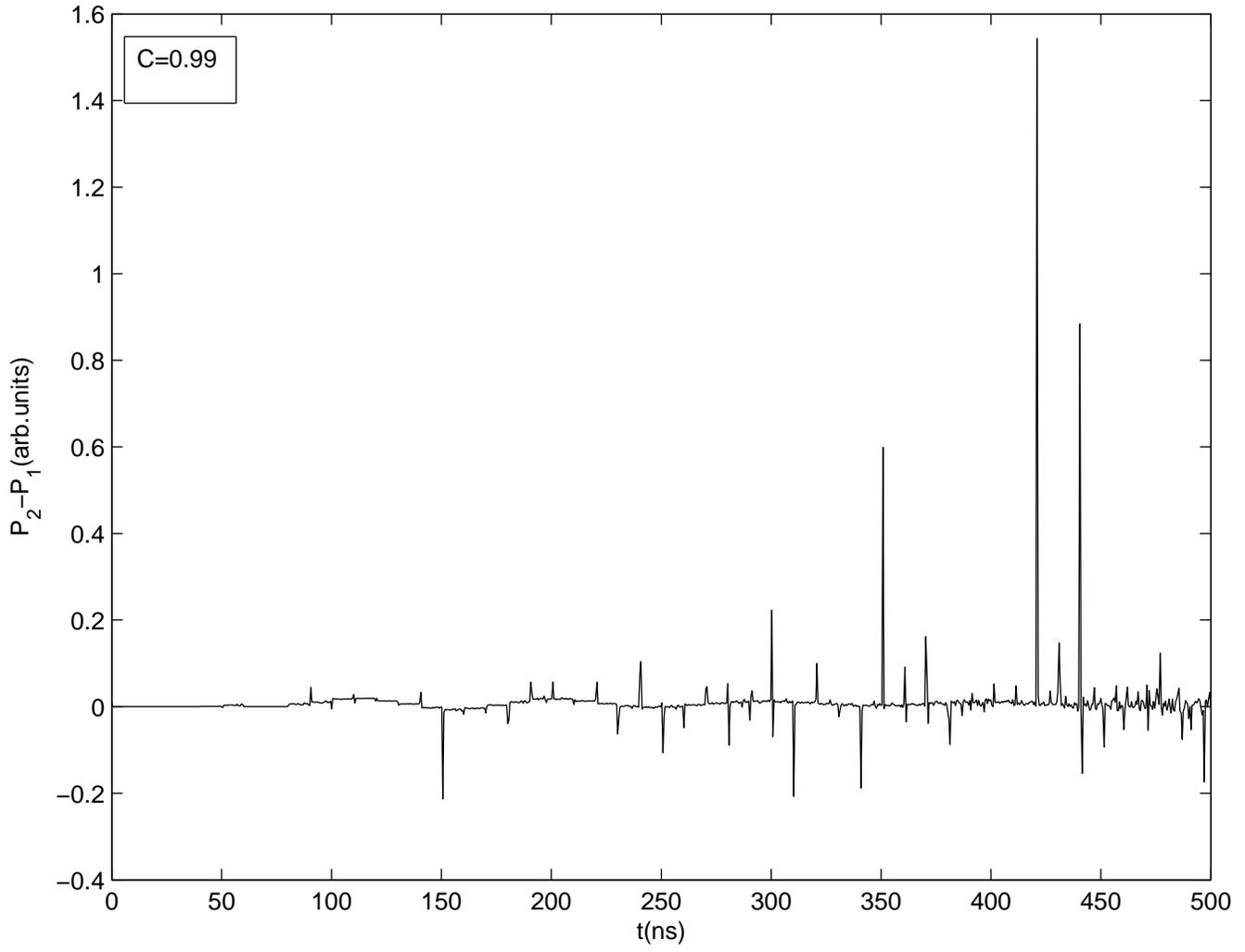
\noindent FIG.11.(a). Error  $P_{2}-P_{1}$ dynamics between two unidirectionally coupled lasers with noise terms included. The parameters are as for figure 3;(b) Error $P_{3}-P_{1}$ dynamics between cascaded unidirectionally coupled lasers, Eqs.(1-4) and (15-16) with noise terms included for $k_{5}=0.11,K_{1}=0.44,$ the other parameters are as for figure 3;(c) Error $P_{2}-P_{1}$  dynamics between two biidirectionally coupled lasers with noise terms included for parameters as for figure 6.\\

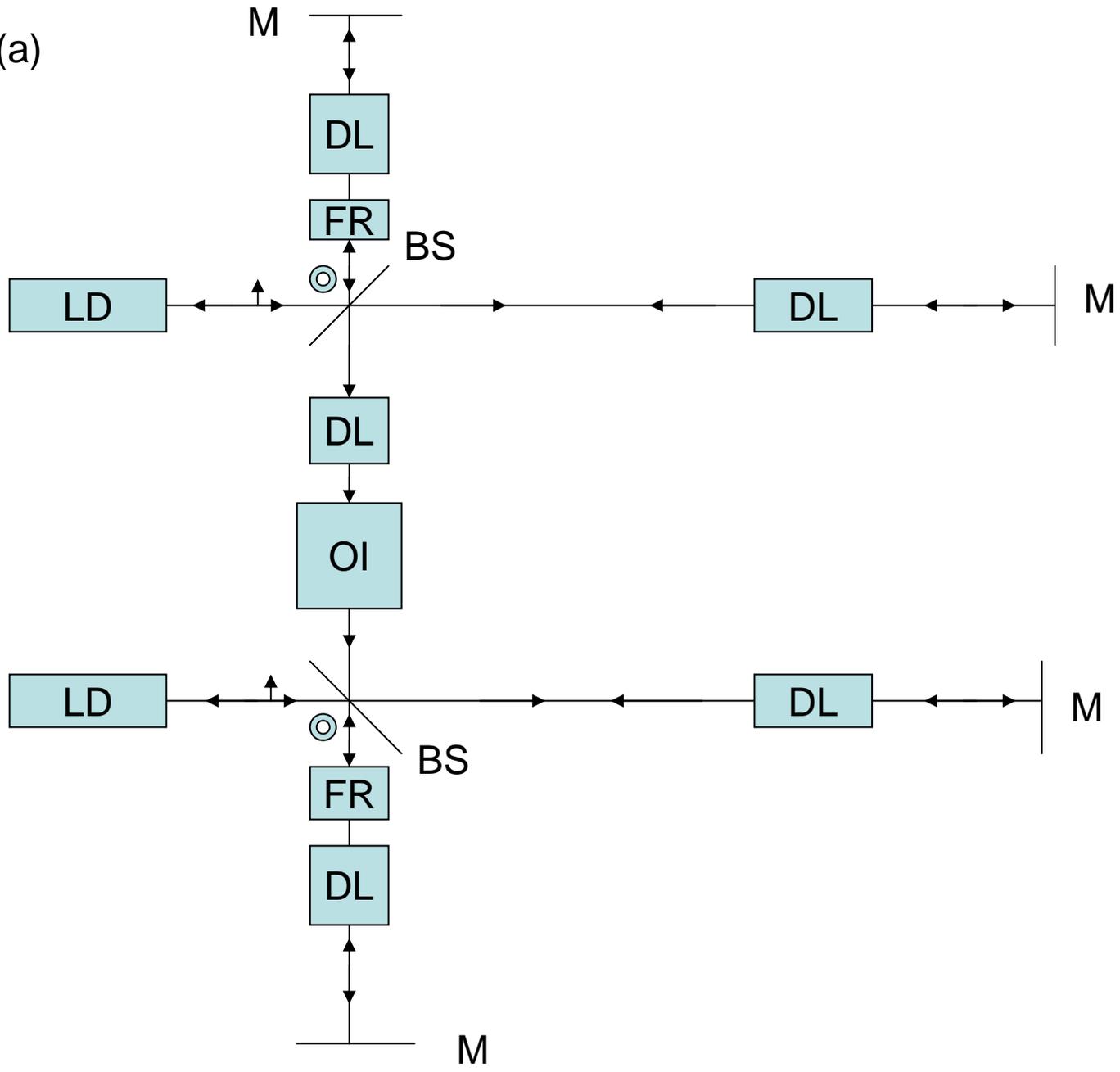
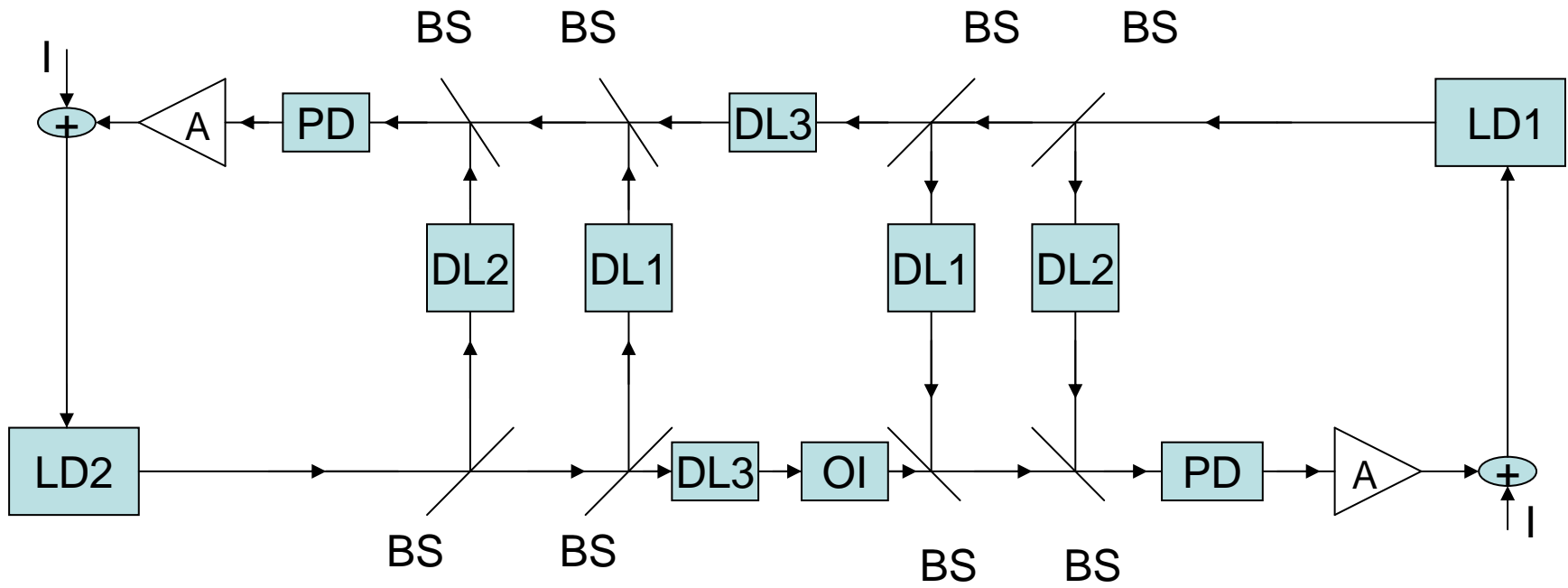
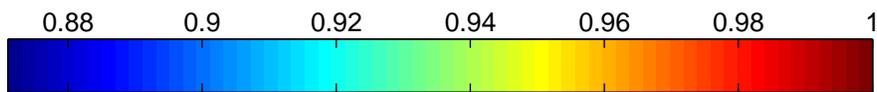
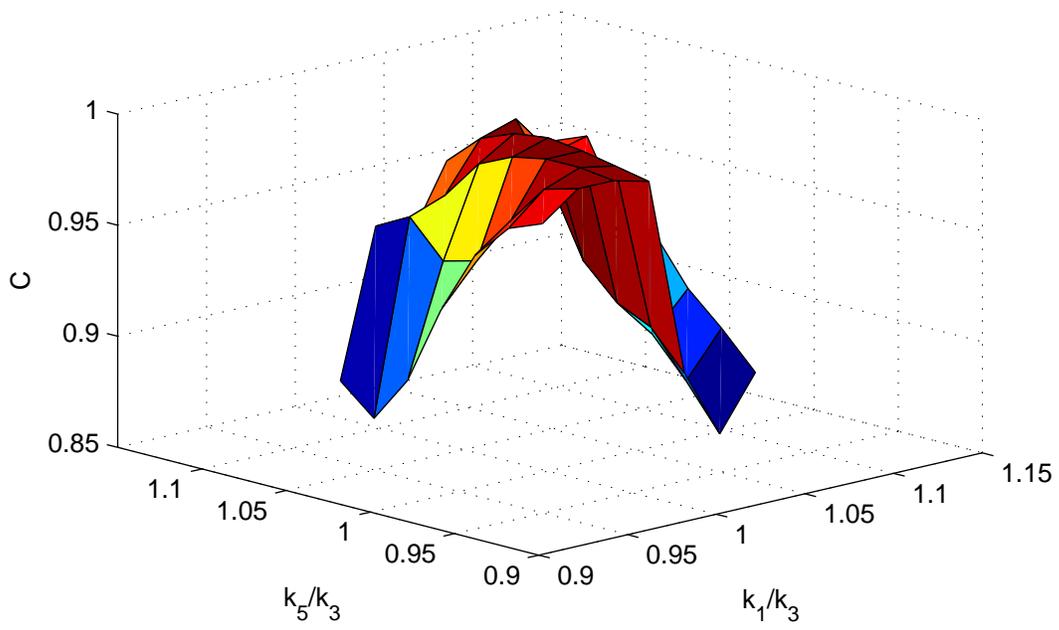
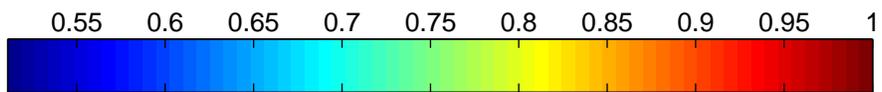
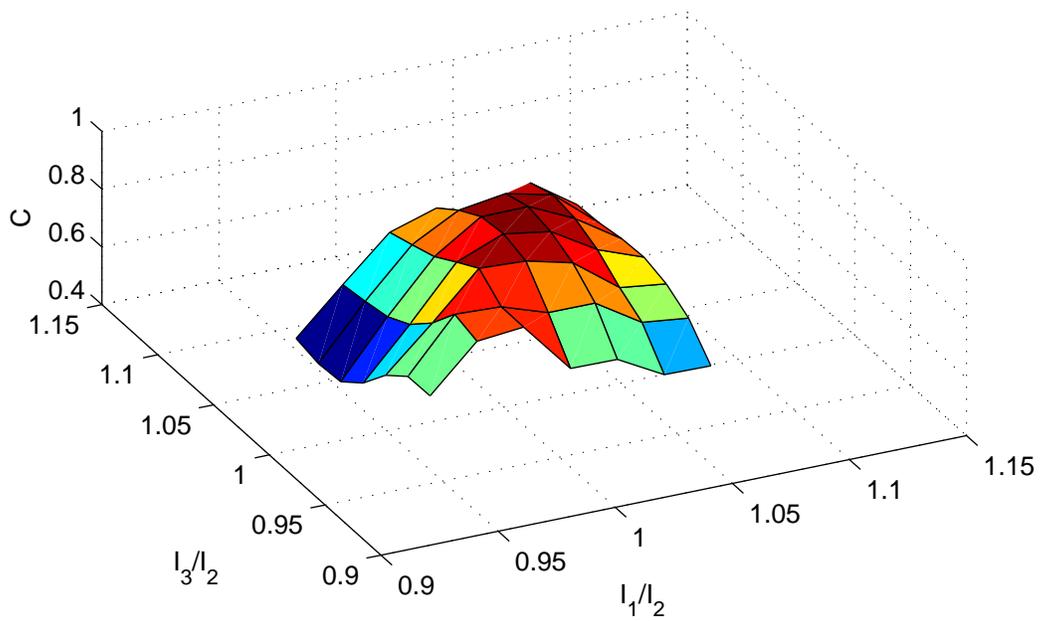
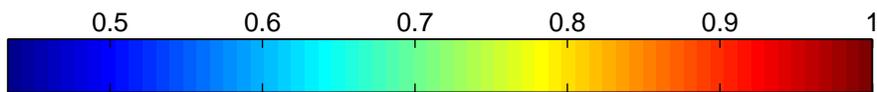
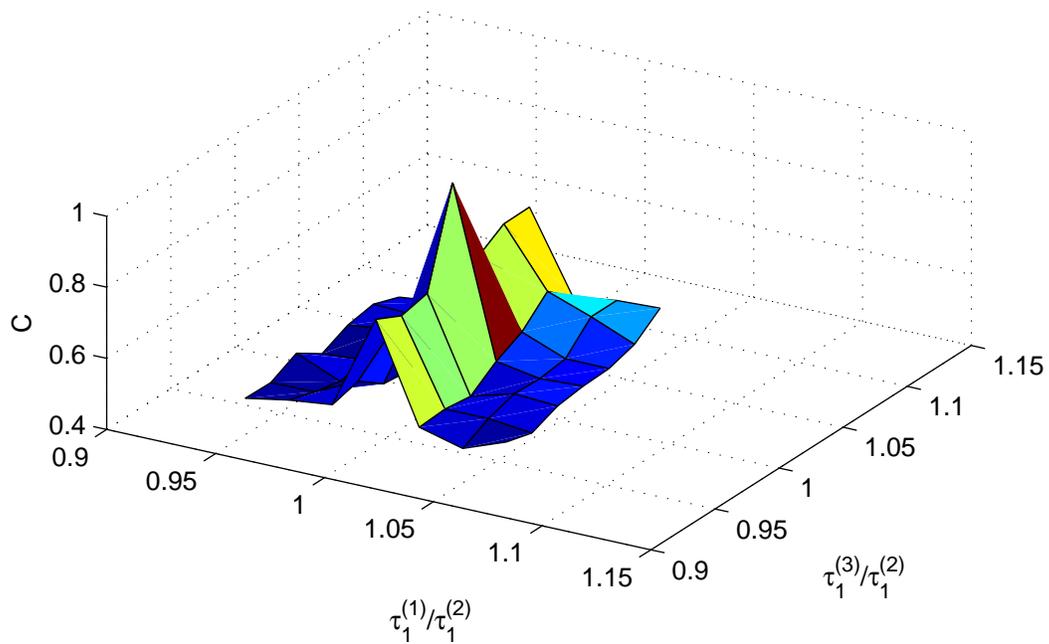
\end{document}